\documentclass[prl,twocolumn,nofootinbib,superscriptaddress]{revtex4}
\usepackage{amsmath,amssymb,epsfig,bbold,bm,mathrsfs,feynmp,slashed,color,graphicx,mathtools}
\usepackage[colorlinks=true,linkcolor=blue,citecolor=blue,urlcolor=blue]{hyperref}
\newcommand{\bea}{\begin{eqnarray}}
\newcommand{\eea}{\end{eqnarray}}

\newcommand{\De}{\Delta}
\newcommand{\de}{\delta}

\newcommand{\beq}{\begin{equation}}
\newcommand{\eeq}{\end{equation}}
\newcommand{\ba}{\begin{array}}
\newcommand{\ea}{\end{array}}
\newcommand{\bi}{\begin{itemize}}  
\newcommand{\ei}{\end{itemize}}
\newcommand{\ben}{\begin{enumerate}} 
\newcommand{\een}{\end{enumerate}}
\newcommand{\bc}{\begin{center}}
\newcommand{\ec}{\end{center}}
\newcommand{\p}{\partial}

\newcommand\eqn[1]{(\ref{#1})}      
\newcommand\Eqn[1]{Eq.~(\ref{#1})}  

\newcommand{\MeV}{{\rm MeV}}

\newcommand{\chid}{\raisebox{0.4ex}{$\chi$}_\delta}
\newcommand{\mud}{\mu_\delta}
\newcommand{\Nd}{N_\delta}

\usepackage[normalem]{ulem}  

\def\XXint#1#2#3{{\setbox0=\hbox{$#1{#2#3}{\int}$}
     \vcenter{\hbox{$#2#3$}}\kern-.5\wd0}}

\definecolor{red}{rgb}{0.8,0,0}
\definecolor{violet}{rgb}{0.4,0,0.4}
\definecolor{green}{rgb}{0,0.5,0.0}
\definecolor{navy}{rgb}{0.0,0.0,0.6}
\definecolor{orange}{rgb}{0.8,0.2,0.0}

\begin{document}
\title{
Color superconductivity and charge neutrality
in Yukawa theory
}

\author{Mark G. Alford}
\affiliation{Physics Department, Washington University, St.~Louis, MO~63130, USA}

\author{Kamal Pangeni}
\affiliation{Physics Department, Washington University, St.~Louis, MO~63130, USA}

\author{Andreas Windisch}
\affiliation{Physics Department, Washington University, St.~Louis, MO~63130, USA}

\begin{abstract}
It is generally believed that when Cooper pairing occurs between two
different species of fermions, their Fermi surfaces become locked together
so that the resultant state remains
``neutral'', with equal number densities of the two species, even
when subjected to a chemical potential that couples to the
difference in number densities.
This belief is based on mean-field calculations in
models with a zero-range interaction, where the 
anomalous self-energy is independent of energy and momentum.
Following up on an early report of a deviation from neutrality
in a Dyson-Schwinger calculation of color-flavor-locked quark matter,
we investigate the neutrality of a two-species
condensate using a Yukawa model which has a finite-range interaction.
In a mean field calculation
we obtain the full energy-momentum dependence of the self energy
and find that the energy dependence leads to a population
imbalance in the Cooper-paired phase when it is stressed by
a species-dependent chemical potential.
This gives some support to the suggestion
that the color-flavor-locked phase of
quark matter might not be an insulator.
\end{abstract}

\maketitle

\medskip\noindent {\it 1. Introduction.}
In a system containing a high density of fermions of two different
species, there may be Cooper pairing between the two species.
This situation arises generically in quark matter, where Cooper pairing
of quarks is most energetically favorable in the flavor-antisymmetric
channel \cite{Bailin:1983bm,Alford:2007xm,Alford:2001dt}, and can also
occur in cold atomic gases \cite{RevModPhys.80.1215}.
It is generally believed that in a BCS condensate with cross-species
pairing, the Fermi momenta of the two species are locked to a
common value, so that even in the presence of a chemical potential 
$\mud$ that
would favor one species over the other, the number densities remain
equal. The charge imbalance $\Nd\equiv N_1-N_2$ remains zero
for a range of $\mud$, so, for example, the
charge imbalance susceptibility, $d\Nd/d\mud$ at $\mud=0$, is zero.

One example where this arises is the color-flavor-locked (CFL)
phase of dense quark matter \cite{Alford:1998mk}. The
CFL phase is believed to be an insulator, with no electrons present,
because BCS pairing between quarks of different flavors
locks the Fermi momenta of the three flavors to the same value, 
ensuring that the quark population remains electrically 
neutral even in the presence of an electrostatic potential that would
favor up quarks over down and strange quarks. Thus the electron density
in CFL quark matter remains zero.

This rigid locking of the Fermi momenta
has been demonstrated for NJL-type models \cite{Rajagopal:2000ff},
where there is a contact interaction between the fermions,
so the fermion self energy, including  
the fermion-number-violating (``anomalous'')
component that arises from pairing,
is independent of the energy and momentum.
However, a study of the CFL phase using the Dyson-Schwinger approach
\cite{Nickel:2008ef}, where the anomalous
self-energy is energy and momentum dependent, found that there were
electrons in the CFL phase. The authors of Ref.~\cite{Nickel:2008ef}
suggest that this arises from
the energy dependence of the anomalous self-energy.

In this paper we show explicitly, in a mean-field treatment
of a model where the fermions interact via
a Yukawa boson, that the energy dependence of
the anomalous self-energy $\Delta$ leads to a 
non-zero charge imbalance susceptibility via a 
factor of $\p\De/\p k_4$ in the relevant integral.
The form factor of 
our interaction is the free boson propagator, not including
any in-medium effects on the boson self-energy. However, this
simple model is sufficient to give an energy and
momentum dependent anomalous self-energy.

The single-flavor Yukawa model was studied previously
by Pisarski and Rischke \cite{Pisarski:1999av}, but they 
neglected the energy-momentum
dependence of the anomalous self energy, arguing that this was valid
at strong coupling.

In very recent work, Sedrakian et~al.~\cite{Sedrakian:2017qpg} studied
a two-flavor model interacting via $\sigma$ and $\pi$ mesons. They
calculated the energy dependence (though not momentum dependence) of the anomalous self energy, and also included in-medium corrections to the boson self energy. They did not calculate the charge imbalance.

\medskip\noindent {\it 2. Yukawa Theory.}
%
Our version of the Yukawa model contains a massless
fermion $\psi$  that comes in two flavors and,
to allow pairing in the $J^P=0^+$ channel, two ``tastes'' which
would become colors in a full QCD treatment.
The fermions have a Yukawa coupling of strength 
$g$ with a Yukawa boson of mass $m$.
The Lagrangian density is
\begin{equation}
\mathcal{L}=\bar{\psi}(i \gamma^\mu \partial_\mu+\gamma^0\mu)\psi+\frac{1}{2}\partial_\mu \phi \partial^\mu\phi-\frac{1}{2}m^2\phi^2-g\bar{\psi}\psi\phi \ .
\end{equation}
The coupling to the Yukawa boson breaks chiral symmetry, so
the internal symmetry group  is 
\beq
\dfrac{SU(2)_{\rm flavor}}{Z_2}\times\dfrac{SU(2)_{\rm taste}}{Z_2} \times
U(1)_{\rm B} \ .
\eeq 
To probe the charge balance of the system we couple the two flavors
to separate chemical potentials
$\mu_1=\mu+\mud$ and $\mu_2=\mu-\mud$. 
We study Cooper pairing in the channel that is a singlet (and
therefore antisymmetric) in flavor, taste, and spin, since this
channel is known to dominate Cooper pairing
in NJL models and in weakly coupled QCD \cite{alford2001color}.
This leaves the flavor and taste symmetries unbroken, and breaks
$U(1)_{\rm B}\to Z_2$.
The free energy of this model in the mean field approximation 
is (see Chapter 5 in Ref.~\cite{Schmitt:2014eka})
\beq
\Omega =\frac{-T}{2 V}\sum_{k}\ln\det\mathcal{S}^{-1}(k)+\frac{8g^2T^2}{V^2} \sum_{q,k}D(k-q)f(q)f(k)\ ,
\label{eq:free_energy1}
\eeq
where $\mathcal{S}^{-1}(k)$ is the inverse fermion propagator, 
the boson propagator is $D(k) = 1/(k^2 + m^2)$,
and $f(q)$ is related to the anomalous self-energy in momentum space
\begin{equation}
\Delta(k)=\frac{g^2T}{V}\sum_{q}D(k-q)f(q) \ .
\label{eq:gap_def}
\end{equation}
The inverse fermion propagator in Nambu-Gor'kov space is
\begin{equation}
\mathcal{S}^{-1} (k)= 
\left(\begin{array}{cc}
S^{-1}_+(k) & T^{-1}_-(k) \\
T^{-1}_+(k)& S^{-1}_-(k)
\end{array}\right)\ ,
\label{eq:inv_prop}
\end{equation}
where the terms on the diagonal are given by
\begin{equation}
S^{-1}_\pm(k)=(\gamma^0(i k_4\pm\mu)-\gamma^ik_i)\otimes \mathbb{1}_f \otimes \mathbb{1}_t\pm\gamma^0\mud \otimes \sigma_{3_f}\otimes \mathbb{1}_t
\end{equation}
and the off-diagonal terms by
\begin{equation}
T^{-1}_\pm(k)=\pm\Delta(k)\otimes \sigma_{2_f}\otimes\sigma_{2_t}\ .
\end{equation}
In each entry, the first factor in the tensor product lives in Dirac space, 
the second factor lives in two dimensional flavor space and the 
third factor lives in two-dimensional taste space. 
The possibility of Cooper pairing is incorporated by the off-diagonal 
(``anomalous'') terms,
which represent the violation of quark number symmetry via the condensate,
allowing a quark to evolve into an anti-quark. The inverse propagator has 
eight distinct eigenvalues, each of which is 4-fold degenerate.
Since the determinant is the product of the eigenvalues, 
we can use this in \Eqn{eq:free_energy1} to obtain the free energy 
\begin{align}
\Omega=& -\frac{2T}{V}\sum_k\log(X) +\frac{8g^2T^2}{V^2} \sum_{q,k}D(k-q)f(q)f(k), \label{eq:free_energy} \\
X\equiv& \prod_{s_i}
 \left(i k_4+s_1 \sqrt{\Delta(k) ^2+(|\vec{k}|+s_2\mu )^2}+s_3 \mud\right)\ ,
\nonumber
\end{align}
where each of the three $s_i$ varies over $\pm 1$ in the product, yielding
8 factors altogether.
Minimizing the free energy 
\eqn{eq:free_energy} with respect to $f(k)$ 
gives the gap equation
\beq
\Delta(k)= \frac{g^2T}{V}\sum_q D(k-q) W(q)\ , 
\label{eq:gap_equation}
\eeq
where we have defined
\beq
W(q) \equiv \frac{1}{4}\sum_{t_1, t_2}\frac{\Delta(q) }{\Delta(q)^2+(|\vec{q}|+t_1\mu )^2+(q_4+t_2i \mud )^2}\ ,
\label{eq:def_W}
\eeq 
where each of the two $t_i$ varies over $\pm 1$ in the sum.
Comparing this with \eqn{eq:gap_def}, we find that at the minimum of the free
energy $f(k)=W(k)$. Using this in \eqn{eq:free_energy}, 
we find the free energy of the mean-field ground state,
\beq
\Omega= -\frac{2T}{V}\sum_k\log(X) +\frac{8T}{V} \sum_{k}\Delta(k)W(k) \ .
 \label{eq:free_energy_min} \\
\eeq
We emphasize that this expression is only valid when $\De(k)$ is a 
solution of the gap equation.

From now on, we work in the zero-temperature limit.

\medskip\noindent {\it 3. Charge imbalance susceptibility.}
We now briefly review the argument that the charge imbalance susceptibility
is zero in NJL models, and describe why it is nonzero
in a Yukawa model. The NJL model is the limit of the Yukawa model
where $D(k-q) \rightarrow 1/m^2$. A momentum cutoff $\Lambda$ is
then required, but the $k_4$ integral can be left unregulated.
The anomalous self-energy $\Delta$
is then independent of energy and momentum, and the gap equation 
\eqn{eq:gap_equation} becomes
\begin{align}
\Delta=\frac{g^2}{2m^2} \!\int \frac{d^4q}{(2\pi)^4}
\biggl( & \frac{\Delta}{\Delta^2+(|\vec{q}|-\mu )^2+(q_4+i \mud )^2} \nonumber \\
  + & \frac{ \Delta }{\Delta ^2+(|\vec{q}|+\mu )^2+(q_4+i \mud )^2} \biggr)\ , 
  \label{Eq:gap_equation_NJL}
\end{align}
where the $q_4$ integration contour can be closed in the upper 
or lower half-plane, since the poles come in complex conjugate pairs
with opposite sign residues.
Since the residues of the poles are independent of $\mud$,
the gap equation is independent of  $\mud$ as long as one can eliminate
$\mud$ by performing a change of integration variable $q_4^\prime=q_4-i\mud$
without moving any
poles from the upper half plane to the lower half plane and vice versa.

The poles of the integrand in the gap equation (\ref{Eq:gap_equation_NJL}) are 
\beq
\label{eq:poles}
q_4=i\left(-\mud\pm\sqrt{\Delta^2+(|\vec{q}|\pm\mu)^2}\right) \ .
\eeq
We can see that the poles with a ``+" sign in front of the square-root 
move toward the lower half plane from the upper half plane as we 
increase $\mud$, and they first cross the real axis when $\mud=\Delta$ 
(with $|\vec{q}|=\mu$). Therefore, as long as $\mud<\Delta$, 
the gap equation and hence the anomalous self-energy $\Delta$ 
and the free energy of the paired state $\Omega_{\rm BCS}$
are independent of $\mud$ and the charge imbalance $\Nd$ and all its
derivatives vanish  \cite{Rajagopal:2000ff}. 

For a Yukawa interaction the gap equation 
(\ref{eq:gap_equation},\ref{eq:def_W}) is
\beq
\Delta(k)= \frac{g^2}{4} \!\!\int\!\!\frac{d^4q}{(2\pi)^4} \frac{1}{(k_4-q_4)^2+(\vec{k}-\vec{q}\,)^2+m^2}W(q) \ .
\eeq
Unlike the NJL case, the presence of a scalar propagator in the gap equation 
results in an energy- and momentum dependent anomalous self-energy. 
Because the anomalous self-energy and the scalar propagator have $q_4$ 
dependence, a shift in $q_4$ by $q_4^\prime=q_4-i\mud$ does not 
result in a $\mud$ independent gap equation. 
This raises the possibility that $\Nd$ and its derivatives may no longer 
be zero. Below we will describe explicit calculations that 
find that this is indeed the case.

\medskip\noindent {\it 4. Number Density and Susceptibility.}
The charge imbalance $\Nd=-d\Omega/d\mud$
only receives a contribution from the first term in
the free energy (\ref{eq:free_energy}), since the second term only
depends on $\mud$ via $\De(k)$, and in the ground state
$\de \Omega / \de \De(k)=0$. Thus
\begin{equation}
\Nd = -\int\frac{d^4k}{(2 \pi)^4}\frac{1}{X}\frac{\partial X}{\partial \mud},
\end{equation}
and the charge imbalance susceptibility $\chid\equiv d\Nd/d\mud|_{\mud=0}$ is
\begin{equation}
\begin{split}
\chid &=\frac{1}{\pi^2}\int dk_4 \ d^3\vec{k}\left(2 k_4^2(U_+^2+U_-^2)-U_+-U_-\right), \label{eq:number_density_derivative} \\
U_\pm &\equiv\frac{1}{\Delta(k) ^2+(|\vec{k}|\pm\mu )^2+k_4^2}.
\end{split}
\end{equation}
%
%
Integrating by parts, we find
\begin{equation}
\chid=\frac{4}{\pi^3}\int dk_4 \ dk k^2\left(\frac{\partial U_+}{\partial \Delta(k)}+\frac{\partial U_-}{\partial \Delta(k)}\right) k_4\frac{\partial \Delta(k)}{\partial k_4}.
\label{Eq:num_density_slope}
\end{equation}
The factor involving $U_\pm$ is negative for all $(k_4,\vec k\,)$.
In NJL models the factor $\p\De/\p k_4$ is zero, but in models with
more realistic interactions, such as the Yukawa model, we expect
that $k_4\p\De/\p k_4$ will be negative since the
interaction between fermions
effectively weakens at high energy or momentum, so the anomalous self energy
should decrease at high energy. Thus we expect $\chid$ to be positive,
meaning that the charge imbalance grows with
the relevant chemical potential $\mud$.

Our numerical results, presented below, confirm these expectations.

\medskip\noindent {\it 5. Numerical Results.}
To study examples of the result obtained above, we numerically solved
the gap equation \eqn{eq:gap_equation}, including all
energy and momentum dependence, by an
iterative procedure.
We discretized the function $\De(k_4,|\vec{k}|)$ on a
grid in energy-momentum space with cutoffs ${\Lambda_k}_{4}$ in the
range $\sim 10^{6}$ to $10^{10}$\,MeV and $\Lambda_{\vec k}$ in the range
$\sim 10^{4}$ to $10^{6}$\,MeV. Since the Yukawa model is renormalizable, we
obtained cutoff-insensitive results as long as
$(\mu,m)\ll\Lambda_{\vec k}\ll{\Lambda_k}_4$.
The typical $(k_4,\vec k)$ grid sizes were $(512\times 256)$. 
The grid points were generated by using a Gauss-Legendre quadrature
rule. The points and weights were then remapped to achieve
better resolution of regions where the integrand shows strong variation. 
The integration was performed on a graphics processing unit. 

\begin{table}
\begin{tabular}{ccccc}
\hline
                                  & $m$ [MeV]           &       25 &       50 &    75\\
\hline
$\Delta_{\rm spectral}\approx50$ MeV  & $g=$                   & 4.787348 & 4.871645 & 4.969172\\
                                  & $\chid=$ & 2472.1   & 2174.6   & 1919.7\\
\hline
$\Delta_{\rm spectral}\approx75$ MeV  & $g=$                   & 5.322668 & 5.388405 & 5.466438\\
                                  & $\chid=$ & 3028.9   & 2791.1   & 2559.8\\
\hline
$\Delta_{\rm spectral}\approx100$ MeV & $g=$                   & 5.726067 & 5.778862 & 5.841499\\
                                  & $\chid=$  & 3470.5   & 3279.9   & 3077.2\\
\hline
\end{tabular}
\caption{
The charge imbalance susceptibility $\chid = d\Nd/d\mud|_{\mud=0}$ for nine
parameter sets,
with different Yukawa boson masses and spectral gaps,
evaluated at fermion chemical potential $\mu=350$\,MeV.
We find $\chid > 0$ in all cases.
}
\label{tab:parameter_sets}
\end{table}

We studied nine different parameter sets, as shown
in Table \ref{tab:parameter_sets}. 
Each row shows a set of theories
with different masses for the Yukawa boson, but with the coupling tuned
to the value shown
so that the spectral gap, which is a physically measurable quantity,
has the same value in all three cases.
We find that the charge imbalance susceptibility is non-zero
and positive, as
our calculations above led us to expect.

\begin{figure}[hbt] 
\includegraphics[width=\hsize]{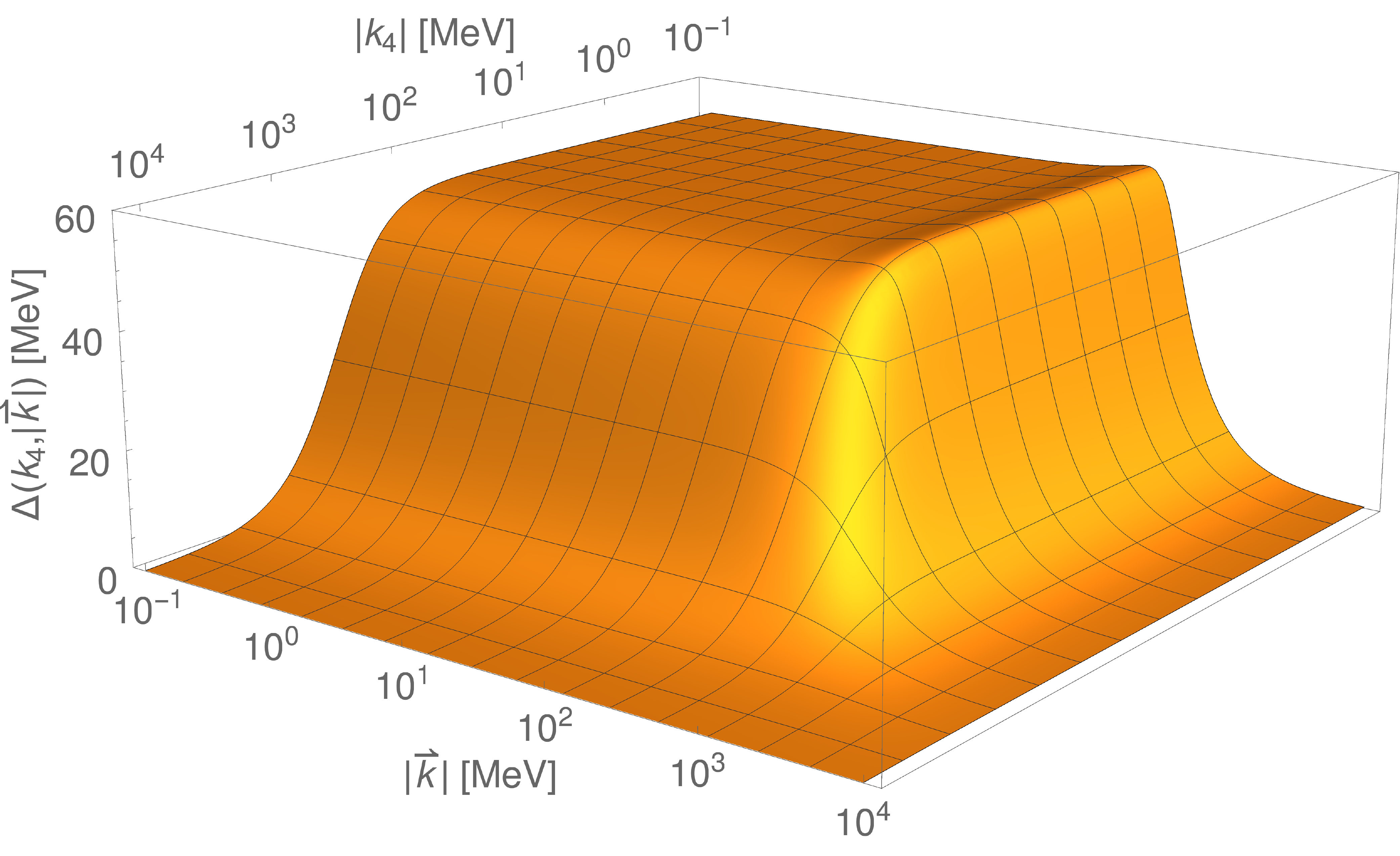}
\caption{The full energy and momentum dependence
of the anomalous fermion self-energy in Yukawa theory
with $m=50$ MeV and $g=4.872$.}
\label{fig:gap_3d_g4871_m50}
\end{figure}

\begin{figure}[hbt] 
\includegraphics[width=0.8\hsize]{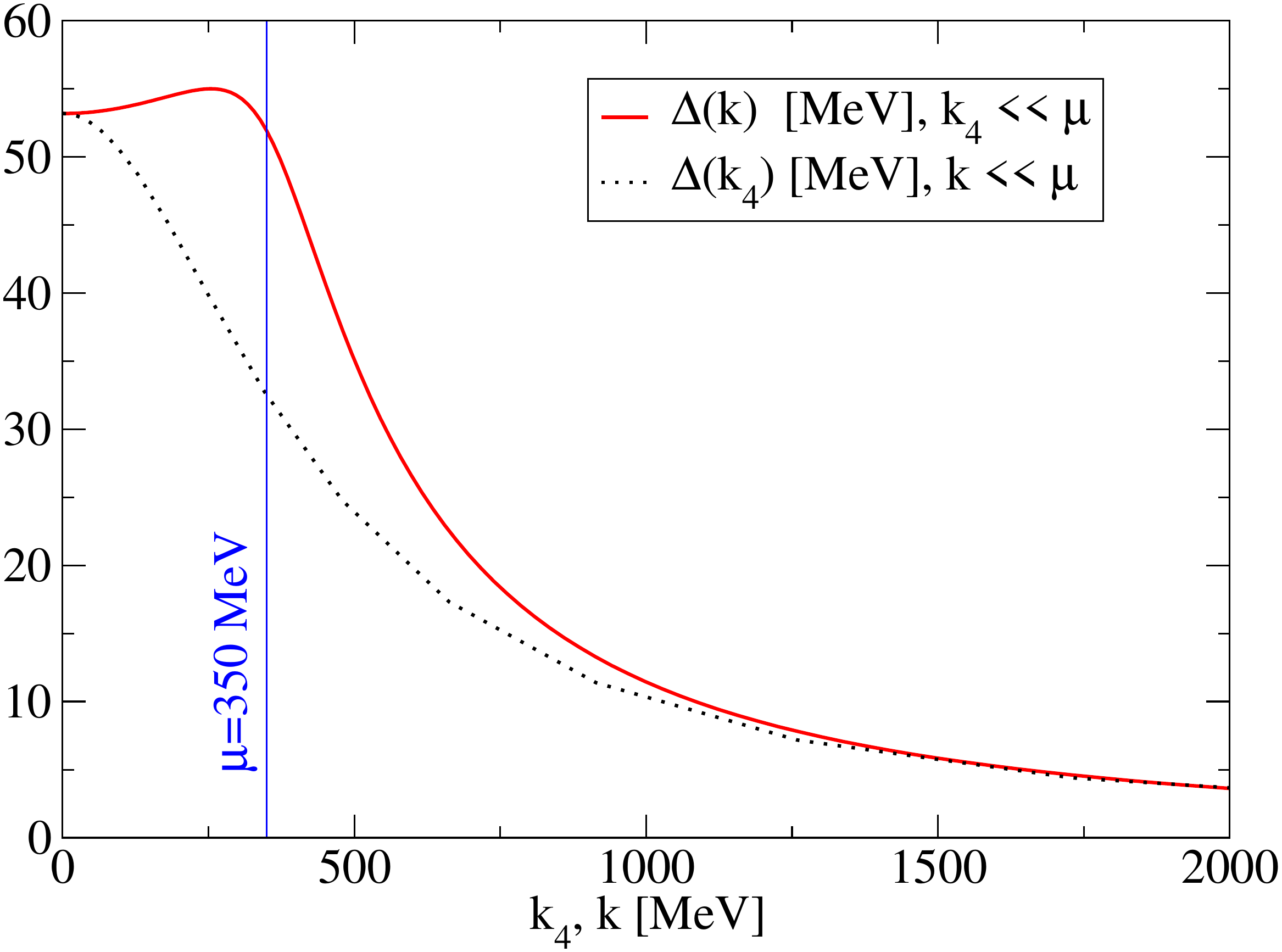}
\caption{Slices through the anomalous self-energy $\De(k_4,k)$ of
Fig.~\ref{fig:gap_3d_g4871_m50}, showing that as a function of energy
$\De$ is monotonically decreasing (dotted curve), but as a function of 
momentum it rises to a maximum around $k=\mu$ and then decreases
(solid curve).}
\label{fig:gap_slices}
\end{figure}

In Fig.~\ref{fig:gap_3d_g4871_m50} we show the
full energy-momentum dependence of the anomalous self-energy $\De$ for
the Yukawa theory with 
$g=4.871645$ and $m=50\,\MeV$, evaluated at $\mu=350$\,MeV.
This case shows the same qualitative properties that we see
for all parameter values that we studied (Table \ref{tab:parameter_sets}).
The anomalous self energy is symmetric in $k_4$, and 
decreases
monotonically with $|k_4|$, so $k_4 d\De/dk_4$ is always
negative, implying via Eq.~\ref{Eq:num_density_slope} that $\chid>0$.
At large $k_4$, $\De\propto 1/k_4^2$.
At large 3-momentum  $|\vec k|$ the self energy
drops off roughly as $1/|\vec k|^{1.7}$, however the behavior is
not monotonic for all  $|\vec k|$. As
$|\vec k|$ rises from zero the anomalous self energy
grows until at reaches a maximum at $|\vec k|\approx\mu$, after
which it decreases monotonically, see Fig.~\ref{fig:gap_slices}.

\begin{figure}[hbt] 
\includegraphics[width=\hsize]{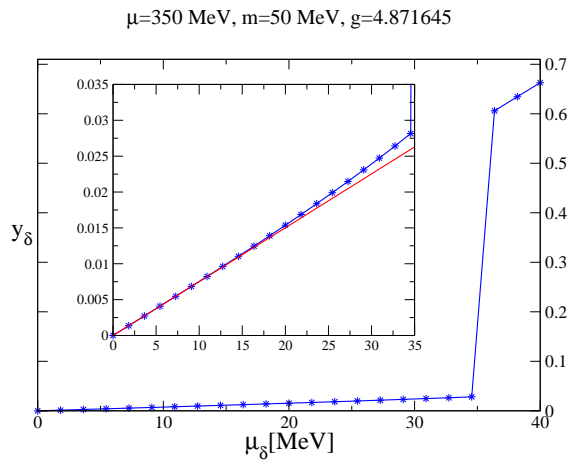}
\caption{Fractional difference $y_\de$ in
the number densities of the two species as a function of the
imbalance potential $\mud$.
The dashed line in the inset plot is the linear approximation
$\Nd=\chid\mud$ based on the charge imbalance susceptibility
$\chid$ (Eqs.~\ref{eq:number_density_derivative},\ref{Eq:num_density_slope}).}
\label{fig:numberdensity}
\end{figure}

Fig.~\ref{fig:numberdensity} shows the fractional
charge imbalance $y_\de = 2\Nd/(N_1+N_2)$,
as a function of the charge imbalance chemical potential $\mud$. 
In an NJL model, where the gap parameter is energy-independent
and the Fermi surfaces are locked by pairing,
it would remain exactly zero until $\mud$ reaches 
$\De_{\rm spectral}/\sqrt{2}$. This is
the Chandrasekhar-Clogston limit 
\cite{chandrasekhar1962note,clogston1962upper} where there is 
a first order transition from a paired state to the unpaired state, 
so $\Nd$ jumps up to around the value expected for free fermions,
$y_\delta$=$2\tilde{\mu}_\delta(3+\tilde{\mu}_\delta^2)/(1+3\tilde{\mu}_\delta^2)$ where $\tilde{\mu}_\delta=\mu_\delta/\mu$.

The solid curve with dots in Fig.~\ref{fig:numberdensity} shows
that in the Yukawa model $\Nd$ does not remain zero, but rises
slowly in response to $\mud$. The inset shows how our numerical calculation
of $\Nd$ agrees with the extrapolation $\Nd \approx \chid\mud$ (solid red
line) from  Eq.~\ref{Eq:num_density_slope}.

\medskip\noindent {\it 6. Conclusions and Discussion.}
%
We have shown that, in a mean field calculation, Cooper pairing between
two different species does not guarantee exact equality in the number
densities of the two species. This marks a qualitative difference
from earlier calculations using
NJL models which found that the charge imbalance was zero
(the Fermi surfaces were ``locked together'') as long as the
system remained in the paired phase.

Our result confirms and 
clarifies the suggestive results of Ref.~\cite{Nickel:2008ef},
which reported such an imbalance in a Schwinger-Dyson
treatment of CFL quark matter.
We studied a simpler system where we could clearly identify
the energy dependence of the self energy as an essential factor
in creating the charge imbalance: the integrand for the
charge imbalance susceptibility 
contains a factor of $k_4\p\De/\p k_4$ (Eq.~\ref{Eq:num_density_slope}).
Since the anomalous self-energy $\De$ drops monotonically as a function
of energy, this means that the susceptibility is generically non-zero.
This implies that the charge imbalance itself is generically non-zero, 
although it can be tuned to zero
by a specific choice of the chemical potential $\mud$
that couples to the difference in number densities.
We cannot rule out the possibility
that our result is an artefact of the mean field approximation, but
either way we must conclude that the previous mean field NJL results 
are not a reliable guide to the
neutrality properties of more realistic theories.

Our result and that of  Ref.~\cite{Nickel:2008ef}
have potentially major implications for the phenomenology of
quark matter in neutron stars. Previous mean field calculations
in NJL models led to the prediction that the
CFL phase of quark matter should be an electrical insulator because
the pairing between the different quark flavors was thought
to ensure that the quark matter contained equal numbers of all
three flavors, so there are no electrons present in neutral CFL
matter. If this prediction
is incorrect, then the phenomenology of the CFL phase, and the signatures
by which it might manifest itself in neutron star observations, will
be significantly affected. A neutralizing population of electrons
would, for example, dominate the specific heat and thermal
conductivity and powerfully 
resist the motion of magnetic field lines \cite{Rajagopal:2000ff}.

The fact that the charge imbalance responds to an arbitrarily
small chemical potential $\mud$ also indicates that there should be
some mode based on the quark degrees of freedom that is massless
and carries an $N_1-N_2$ charge. Further
investigation of this mode, 
along with the study of in-medium corrections to
the Yukawa boson propagator, beyond-mean-field effects,
an analogous investigation of fermions
interacting via gluons, and the
phenomenological consequences for quark matter, would be natural
extensions of this work.


\begin{acknowledgments}
This research was partly supported by the U.S. Department of Energy, 
Office of Science, Office of Nuclear Physics under Award No.
\#DE-FG02-05ER41375. AW acknowledges support by the Schr\"odinger Fellowship J~3800-N27 of the Austrian Science Fund (FWF).
\end{acknowledgments}

\bibliography{gap_eq_Yukawa}{}
\bibliographystyle{apsrev4-1}  

\end{document}